\documentclass[useAMS,usenatbib]{mn2e}
\usepackage{epsfig}
\usepackage{amssymb}

\title[Aspherical SN explosions and formation of BHLMXBs]
{Aspherical supernova explosions and formation of compact black hole
low-mass X-ray binaries}
\author[Xiang-Dong Li]{Xiang-Dong
Li\thanks{E-mail:lixd@nju.edu.cn }\\
Department of Astronomy, Nanjing University, Nanjing 210093, China
}
\begin{document}

\date{Accepted . Received ; in original form }

\pagerange{\pageref{firstpage}--\pageref{lastpage}} \pubyear{2007}

\maketitle

\label{firstpage}

\begin{abstract}
It has been suggested that black-hole low-mass X-ray binaries
(BHLMXBs) with short orbital periods may have evolved from BH
binaries with an intermediate-mass secondary, but the donor star
seems to always have higher effective temperatures than measured in
BHLMXBs \citep{just06}. Here we suggest that the secondary star is
originally an intermediate-mass ($\sim 2-5 M_{\sun}$) star, which
loses a large fraction of its mass due to the ejecta impact during
the aspherical SN explosion that produced the BH. The resulted
secondary star could be of low-mass ($\la 1 M_{\sun}$). Magnetic
braking would shrink the binary orbit, drive mass transfer between
the donor and the BH, producing a compact BHLMXB.
\end{abstract}


\section{Introduction}

There exist currently around twenty stellar-mass black hole (BH)
candidates in binary systems \citep{casa06,remi06}. Nine of these
systems are defined as compact BH low-mass X-ray binaries (LMXBs)
with short orbital periods ($\la$ 1 d) and donors of mass $\la
1M_{\odot}$ \citep{lee02,ritt03,pods03}. The short orbital periods
of these BHLMXBs imply that they must have undergone secular orbital
angular momentum loss, since mass transfer from the less massive
donor star to the more massive BH always causes the orbit to widen.
The standard formation scenario, in which the progenitor systems
contains a massive primary and a low-mass secondary initially, faces
several difficulties, as summarized in \citet{just06} and
\citet{ivan06}. (1) The progenitor binaries with extreme mass ratios
($>20$) are very difficult to form, considering the fact that
massive stars tend to have binary companions with similar masses
\citep{pins06}; (2) the secondary star, because of its low mass, may
not have enough energy to eject the envelope of the BH progenitor
during the common envelope evolution phase, unless a significant
fraction of the envelope has been previously lost through a very
efficient stellar wind; (3) the binary is likely to be disrupted
during the supernova (SN) explosion that produced the BH.

There have been quite a few alternative scenarios suggested for the
formation of compact BHLMXBs. \citet{eggl86} suggested that the
progenitor of a BHLMXB is a triple star in which a massive close
binary is accompanied at a large distance by a late dwarf. After the
evolution of the close binary into an ordinary X-ray binary, the
compact object is engulfed by its expanding massive companion, and
spirals in to settle at its center. The resulting Thorne-\.{Z}ytkow
object (TZO) gradually expands until it attains the size of the
late-dwarf orbit. Then a second spiral-in phase ensues, leading to
the formation of a low-mass close binary. But it is difficult for
this scenario to explain the spatial distribution and space
velocities of LMXBs. \citet{pods95} proposed that during the
evolution of TZOs the central neutron star (NS) may be converted
into a BH by accretion, and part of the envelope may collapse into a
massive disc, which may become gravitationally unstable and lead to
formation of low-mass stars or planets. The efficiency of NS
accretion in this case is however, highly uncertain. \citet{ivan06}
suggested that a subset of short-period BHLMXBs could be powered by
mass transfer from pre-main-sequence donors, although they suffer
from the short lifetime problem ($\sim 10^7$ yr). \citet{pods03} and
\citet{just06} assumed that the secondary star is initially an
intermediate-mass star, which is more likely to survive the common
envelope evolution to form an intermediate-mass X-ray binary (IMXB).
Along with mass transfer between the secondary and the BH, the
secondary's mass decreases to be $\la 1 M_{\sun}$, and the binary
becomes an LMXB with long lifetime ($\sim 10^9-10^{10}$ yr). To
maintain long-term orbital shrinkage, the donor stars are further
assumed to be Ap and Bp stars so that magnetic braking can work
(alternatively, a circumbinary disc could do the same job without
requiring the secondary stars to possess anomalously high magnetic
fields, see \citet{chen06}). However, the calculated effective
temperatures of the donor stars are not compatible with the observed
values, which indicate that the donor masses should be $\la 1
M_{\sun}$ at least at the onset of mass transfer.

Now the situation is that, the formation processes require that the
secondary is likely to be an intermediate-mass star, while apparent
donor spectral classes suggest the donor star would be of low mass
all the way. A plausible solution to this puzzle is that, the
secondary is initially an intermediate-mass star, but loses a
significant part of its mass after the formation of the BH. In this
work we explore the possibility of mass loss from the secondary by
the impact from the aspherical SN ejecta.

\section[]{Mass loss due to SN impact}

BH formation may be associated with a SN explosion. In the framework
of current stellar evolution theories, stars more massive than $\sim
40 M_{\sun}$ collapse to BHs directly with no SN explosions. If the
initial stellar mass is between $\sim 20$ and $\sim 40 M_{\sun}$, a
BH forms in a two-stage process, where the collapse first leads to
the formation of a NS accompanied by a SN, which is subsequently
converted into a BH through accretion from the SN fallback
\citep[e.g.][]{fryer99}.

Observationally, \citet{isra99} and \citet{gonz07} have presented
favorable evidence that the BH in the LMXB Nova Sco 1994 (GRO
J1655$-$40) formed in a SN event. From high-resolution spectrum,
they found that the atmosphere of the companion was enriched by a
factor of $6-10$ in several $\alpha$-process elements (O, Mg, Si and
S), indicating that the compact primary most likely formed in a SN
event of a massive star whose nucleosynthetic products polluted the
secondary, since some of these elements cannot have been produced in
a low-mass secondary \citep[see however][for negative
argument]{foe07}. \citet{gonz06} also found supersolar abundances of
Mg, Al, Ca, Fe, and Ni in the atmosphere of the companion star in
the BHLMXB XTE J1118$+$480, and reached the similar conclusion.
Additional independent evidence that the BH was formed in a SN
explosion comes from the peculiar velocities measured for GRO
J1655$-$40 \citep{mira02} and XTE J1118$+$480 \citep{gual05}.

The SN explosion can influence the companion's structure through
hydrodynamic impact. The impact of the SN ejecta on a nearby
companion may be quite dramatic. The supernova ejecta may either
directly strip material from the companion by direct transfer of
momentum or, evaporate the envelope through the conversion of the
blast kinetic energy into internal heat
\citep{mccl71,suta74a,suta74b,chen74}. \citet{whee75} analytically
estimated the amount of mass lost from the companion as a result of
the inelastic collision and the shock heating. They found that, in
the case of spherical SN explosions, the fraction of lost mass from
the companion depends on the value of the parameter $\Psi$, which is
defined as
\begin{equation}
\Psi=\frac{1}{4}\frac{M_{\rm SN}}{M_{\rm
c}}\frac{R^2}{a_0^2}(\frac{v_{\rm SN}}{v_{\rm es}}-1),
\end{equation}
where $M_{\rm SN}$ is the mass of the SN ejecta, $M_{\rm c}$ the
mass of the companion, $R$ the radius of the companion, $a_0$ the
orbital separation just before the SN, $v_{\rm SN}$ the ejecta
velocity, and $v_{\rm es}$ the escape velocity from the companion,
respectively. Note that $v_{\rm es}$ is weakly dependent on the
position within the regions to be stripped and ablated. It would be
increased by about a factor of $\la 2$ at the bottom of the
envelope. For an $n=3$ polytrope, which is appropriate for a
unevolved star, half of the stellar mass is ejected when $\Psi\sim
2$, and the star is completely destroyed when $\Psi\sim 10$.
Numerical simulations were performed for supernova impacts on both
low-mass main-sequence companions \citep{fryx81,taam84} and low-mass
red giants \citep{livn92}. The most recent high-resolution
hydrodynamic simulations made by \citet{mari00} show that the
analytic estimates by \citet{whee75} do provide ballpark estimates
of the ejected mass for the main-sequence star case when $\Psi<1$.

However, high degree of polarization measured in several SNe
\citep[e.g.][]{wang01,leon02,leon05} strongly suggests that perhaps
most SN explosions are aspherical. SNe associated with gamma-ray
bursts (GRBs) are obviously aspherical, as GRBs are generally
believed to be highly asymmetric phenomena \citep{woos06}. The light
curve and the nebular line features of GRB-SN 1998bw were found to
be in conflict with what is expected from a spherically symmetric
explosion model \citep{mazz01}. \citet{maed02} showed that this
configuration can be obtained in an axisymmetric explosion. In such
an explosion, Fe is mostly ejected at high velocity in a jet along
the polar direction, while nearer the equatorial plane relatively
low-velocity O is mostly ejected. It is now widely believed that
these most energetic SNe, described as ``hypernovae", are bipolar
explosions. Assume that the secondary is impacted by a jetlike SN
debris with a solid angle $\Omega$, Eq.~(1) can be then modified to
be
\begin{equation}
\Psi=\frac{\eta}{4}\frac{M_{\rm SN}}{M_{\rm
c}}\frac{R^2}{a_0^2}(\frac{v_{\rm SN}}{v_{\rm es}}-1),
\end{equation}
where $\eta=4\pi/\Omega$. Note that here $M_{\rm SN}$ and $v_{\rm
SN}$ correspond to the the jetlike component in the SN ejecta. All
hypernova models show that the jet emerges along the rotation axis
of the compact object \citep[e.g.][and references therein]{burr07},
which is implicitly assumed to be perpendicular to the orbital plane
if in binary systems. However, anisotropic SN explosion can lead to
misalignment between the spin and orbital axes. In at least two BH
binaries GRO J1655$-$40 and SAX J1819$-$2525 the observed
relativistic jets appear not to be perpendicular to the orbital
plane \citep[][and references therein]{macc02}. If the jet
directions are indicative of the direction of the spin of the BH,
then the most likely explanation is that the misalignment occurred
during the formation process of the BH, and that subsequent
evolution has not had time to bring about alignment \citep{king05}.

According to the calculations by \citet{maed02} and \citet{maed06}
for SN 1998bw, the opening angle of the ejecta in the polar
direction is around $30\degr$. If we adopt a more conservative value
of $45\degr$, $\Omega\sim 0.3\pi$. Fits to the observational data
suggest that the velocity of the ejecta in SN 1998bw can be high as
a few $10^4$ kms$^{-1}$. For a $4M_{\sun}$ main-sequence secondary
in 1 day orbit, inserting typical values for the parameters in the
above equation, we have
\begin{equation}
\Psi \simeq 2.5 (\frac{\eta}{10})[\frac{(M_{\rm SN}/4M_{\sun})}
    {(M_{\rm c}/4M_{\sun}})][\frac{(R/3R_{\sun})}{(a_0/10R_{\sun})}]^2
    [\frac{(v_{\rm SN}/10^4\,{\rm kms}^{-1})}{(v_{\rm es}/800\,{\rm
    kms}^{-1})}-1],
\end{equation}
suggesting that a large fraction of the stellar mass could be lost.
Note that mass ejection is efficient for narrow systems. This is in
accordance with the fact that the postexplosion binary has to be
close enough to start the mass transfer driven by orbital angular
momentum loss. Population synthesis calculations show that BHXBs
with intermediate-mass ($\sim 2-5 M_{\sun}$) secondary are born with
orbital periods in the range from $\sim 0.5$ day to $\la 5$ days
\citep{pods03}. Finally we emphasize that the above estimate is just
of order of magnitude, and may have substantial errors due to the
uncertainties in the morphology of hypernova explosions and the
interaction of the SN ejecta with the secondary especially in the
case of $\Psi\ga 1$. These issues could only be resolved by future
high-resolution numerical simulations of aspherical SN-secondary
interactions. But Eq.~(3) does suggest that we need to seriously
consider the possibility of efficient mass loss from the secondary
by SN impact.

According to the calculations by \citet{mari00}, immediately after
the impact, the secondary star is puffed up, much like a
pre-main-sequence star. Since the remaining envelope is out of
thermal equilibrium, the luminosity of the remnant will vary
dramatically with a Kelvin-Helmholtz timescale $\sim 10^3-10^4$ yr.
After thermal equilibrium is reestablished, the remnant will return
to the main sequence along a Kelvin-Helmholtz track and then will
continue its evolution at a rate prescribed by its new mass. If now
the stellar mass is $\la 1 M_{\sun}$ and the orbital period $\sim 1$
day, magnetic braking will cause the orbit to shrink, and drive mass
transfer onto the BH, leading to the formation of a compact BHLMXBs.

\section{Discussion}

In this work we argue that BHLMXBs may begin with an
intermediate-mass secondary, and that there could be rapid mass loss
at the birth of the BH accompanied by a SN explosion. In the case of
highly aspherical SN explosions, the ejecta impact could strip and
blown the majority of the secondary's mass, leaving it to be a
low-mass star. If the orbit is close enough, with the help of
magnetic braking, the binary will evolve to a short-period LMXB. The
low-mass donor also makes it possible to account for the observed
cool spectral types. If the secondary loses little mass during the
SN explosion, an IMXB will form, which will ultimately evolve to be
a wide LMXB. Note that the efficient mass loss requires high
asphericity of the SN explosion and fine tuned ejecta direction.
This also means that the occurrence rate is likely to be quite low,
and that only a small fraction of the progenitor binaries may become
compact LMXBs through this channel. But the mass transfer lifetime
of these binaries can be as long as $\sim 10^{10}$ yr, at least one
order of magnitude larger than that of wide LMXBs
\citep[e.g.][]{just06}. Hence the expected populations of long- and
short-period BHLMXBs could still be roughly equal in size.

In the following we briefly discuss possible observational clues to
rapid mass loss during a SN explosion. (1) If the SN ejecta is
contaminated with stripped hydrogen from the secondary star, the SN
may appear as a type IIb or IIc SN if the collapsing star has lost
most of its hydrogen or helium envelope: The spectra undergo a
transformation between a hydrogen-rich type II SN and a helium-rich,
hydrogen-deficient type Ib or a hydrogen, helium-deficient type Ic
SN. (2) After the SN and ejecta impact, the binary (if not
disrupted) is likely to have a significant orbital eccentricity, at
periastron the secondary could overfill its Roche-lobe and transfer
matter to the compact object. The X-ray source 1E 161348$-$5055 in
the SNR RCW 103 \citep{tuo80} might be such an example. Recent
observations with {\em XMM-Newton} showed a strong periodic
modulation at $6.67\pm 0.03$ hr \citep{luc06}. If this period is of
orbital origin, 1E1613 could be a young NS or BH accreting from a
very low-mass companion star, as optical/IR observations suggested
that the possible companion should be less massive than
$0.4\,M_{\sun}$, if it is a normal star \citep{pav04,wan07}. Since
it is extremely difficult for such a low-mass star to survive the
common envelope evolution and SN explosion, the companion star might
originate from an intermediate-mass star which has experienced
strong SN impact. We expect detailed multiwavelength observations of
this source to verify or falsify this ejecta impact predication.

\section*{Acknowledgments}
We are grateful to Dr. Jingsong Deng for helpful discussion about
hypernova simulations, and an anonymous referee for constructive
comments. This work was supported by the National Science Foundation
of China under grant numbers 10573010 and 10221001.

\bsp

\label{lastpage}

\end{document}